\documentclass[prb,twocolumn,superscriptaddress, showpacs]{revtex4-1}

\usepackage{amsfonts}
\usepackage{verbatim}
\usepackage{times}
\usepackage{amsmath, amsthm, amssymb}
\usepackage{graphicx}
\usepackage{color}
\usepackage{natbib}
\usepackage{hyperref}
\hypersetup{
        colorlinks=true,
}

\usepackage{tikz}
\usetikzlibrary{calc}
\usetikzlibrary{backgrounds}
\usetikzlibrary{arrows}
\usetikzlibrary{shapes.arrows}
\usetikzlibrary{decorations.markings}

\newcommand{\be}{\begin{equation} }
\newcommand{\ee}{\end{equation} }
\newcommand{\ba}{\begin{eqnarray} }
\newcommand{\ea}{\end{eqnarray} }

\begin{document}


\title{Magnetic and Superconducting Ordering at LaAlO$_3$/SrTiO$_3$ Interfaces}

\author{Lukasz Fidkowski}
\affiliation{Station Q, Microsoft Research, Santa Barbara, CA 93106-6105, USA}
\author{Hong-Chen Jiang}
\affiliation{Station Q, Microsoft Research, Santa Barbara, CA 93106-6105, USA}
\affiliation{Kavli Institute for Theoretical Physics, University of California, Santa Barbara, CA 93106}
\affiliation{Center for Quantum Information, IIIS, Tsinghua University, Beijing, 100084, China}
\author{Roman M. Lutchyn}
\affiliation{Station Q, Microsoft Research, Santa Barbara, CA 93106-6105, USA}
\author{Chetan Nayak}
\affiliation{Station Q, Microsoft Research, Santa Barbara, CA 93106-6105, USA}
\affiliation{Department of Physics, University of California, Santa Barbara, CA 93106}

\date{compiled \today}

\begin{abstract}
We formulate a model for magnetic and superconducting ordering at LaAlO$_3$/SrTiO$_3$ interfaces
containing both localized magnetic moments and itinerant electrons. Though these both
originate in Ti 3d orbitals, the former may be due to electrons more tightly-bound
to the interface while the latter are extended over several layers. Only the latter
contribute significantly to metallic conduction and superconductivity. In our
model, the interplay between the two types of electrons, which is argued
to be ferromagnetic, combined with strong spin-orbit coupling of the
itinerant electrons, leads to magnetic ordering.  Furthermore, we propose a model for interfacial superconductivity, consisting of random superconducting grains in the bulk STO driven, via coupling to the interface conduction band, towards long-ranged or quasi-long-ranged order.  Most interestingly, the magnetic order and strong spin orbit coupling can lead in this manner to unconventional interfacial superconductivity, yielding a possible realization of Majorana physics.

\end{abstract}

\maketitle


\section{Introduction}
It was recently discovered that,
although LaAlO$_3$ and SrTiO$_3$ are both insulators,
the interface between them is metallic\cite{Ohtomo04}.
Furthermore, the electrons at this interface
have shown a variety of remarkable properties, including
magnetism\cite{Brinkman07,Ariando11,Li11,Bert11}
and superconductivity\cite{Reyren07}. Magnetism and superconductivity
often appear in the phase diagram of strongly-correlated materials,
where they compete. However, it is very unusual for them to occur
simultaneously, which appears to be the case at the
LAO/STO interface \cite{Li11,Bert11, MannhartSchlom}.
A basic question, then, is whether the same electrons are exhibiting
both superconductivity and magnetism or if, instead, there is a precise sense in
which there are two species of electrons -- two different electronic bands, for instance --
one of which is superconducting and the other of which is magnetic.
In the former case, the superconductivity must be exotic, perhaps
$p$-wave superconductivity or a Fulde-Ferrel-Larkin-Ovchinnikov (FFLO)
state~\cite{FF, LO}. This would contradict the conventional wisdom that the interface
electrons are simply exhibiting the superconductivity of doped SrTiO$_3$,
which is presumed to be a phonon-mediated $s$-wave superconductor.

The experiments that reveal interesting magnetic behavior
fall into two classes, those which deduce magnetism from
transport and those which attempt to measure it more directly. The former
include experiments that observe hysteresis in the electrical resistance as a
function of magnetic field \cite{Brinkman07,Ariando11}, which show that
there is a magnetic field-driven first-order phase transition
which has a large effect on the resistance. A magnetic transition
is the most natural hypothesis. These signatures are found up to
temperatures in excess of $200$K. The latter include
a torque magnetometry measurement \cite{Li11} that
shows that a field as small as a few milliTesla leads to a large
magnetization, approximately $0.3 \, \mu_B$ per interface unit cell.
This implies that the system has ferromagnetic domains which become aligned
by even a very small field. This experiment shows that the magnetic moment,
which points in the plane, has an onset temperature that is
at least as high as $40$K and persists below the superconducting
$T_c$. Finally, scanning SQUID magnetometry \cite{Bert11}
finds micron sized ferromagnetic domains in a paramagnetic background.
From their estimates, most of the interfacial electrons which are predicted by
polar catastrophe arguments \cite{Ohtomo04} are paramagnetic.
An order of magnitude smaller number of electrons are in ferromagnetic
regions, and a two orders of magnitude smaller number of electrons are in
superconducting regions. There does not appear to be any correlation
between the magnetic and superconducting regions (unless there is spatial segregation in the $z$ direction), which implies that magnetism in a region does not prevent superconductivity from occurring in that region, counter to conventional wisdom.
Taken together, these experiments constitute strong evidence that
there are ferromagnetic domains at the LAO/STO interface that
strongly affect normal state transport and
are also present in the superconducting state.

Since neither LAO nor STO is magnetic, there is clearly
a puzzle here: what is the cause of (at least local) ferromagnetism
at their interface? It has been suggested that there is a narrow band at
the interface \cite{Pentcheva06} which gives rise to itinerant electron
ferromagnetism. Alternatively, the magnetism may be due to
localized electrons which don't participate in the metallic (or superconducting)
behavior. To make matters even less clear, there is evidence \cite{Caviglia10,Triscone12}
for strong spin-orbit coupling due to the broken inversion symmetry
of the interface \cite{BenShalom} (Rashba spin-orbit coupling), which would ordinarily be
antithetical to a uniform ferromagnetic moment.

An equally vexing problem is how ferromagnetism can coexist
with superconductivity \cite{Dikin11}. Even if there were a sense in which
different electronic bands were becoming ferromagnetic
and superconducting, one would expect the magnetic moments
of the former electrons to destroy superconductivity in the latter.
One possibility is that the system is in an inhomogeneous
superconducting state, e.g. the FFLO state, as suggested in Ref. \onlinecite{Michaeli12}.
Or, one could imagine larger scale inhomogeneity, so that the system
breaks up into domains, some of which are superconducting
while others are ferromagnetic. But the ferromagnetic
moment would be anti-correlated with superconductivity in either
type of inhomogeneous state.
For instance, in the FFLO state, the system forms superconducting stripes,
separated by magnetic ones. This disagrees with the experimental
finding of Ref. \onlinecite{Bert11}. Therefore, it is natural to consider, instead,
a $p$-wave superconducting state, but this begs question
of what the superconducting mechanism is. Presumably,
$p$-wave superconductivity must be due to electron-electron
interactions, rather than the electron-phonon coupling.
Thus, superconductivity at the LAO/STO interface is a puzzle.
It is generally assumed that it is related to superconductivity
in doped STO, but this assumption does not provide any clues
to how it can coexist with ferromagnetism.

In this paper, we present a physical picture for magnetism and
superconductivity at the LAO/STO interface.
Density functional theory calculations show that there are
$t_{2g}$ bands at the interface, corresponding to Ti $3d$ orbitals,
with $d_{xy}$ and $d_{xz,yz}$ symmetry respectively
\cite{Pentcheva06,Popovic}.
We hypothesize that the $d_{xy}$ band forms a band of localized electrons
which accounts for most of the charge required by the
`polar catastrophe' \cite{OkamotoMillis}. According to our picture, the $d_{xz,yz}$
bands form metallic bands of itinerant electrons at the interface.
Coulomb interactions between localized
and itinerant electrons generates a ferromagnetic interaction between
them, thereby leading to a ferromagnetic Kondo model -- but one in which
the itinerant electrons have significant Rashba spin-orbit coupling.
By analyzing a spin-orbit-coupled ferromagnetic Kondo lattice, we argue
that the localized electrons develop magnetic order.
Thus, in our picture, the magnetic moment of the system
is due primarily to localized electrons.

Our picture for superconductivity is the following.
We suppose that there are droplets of local superconductivity in
the STO substrate. If the STO were doped, then these droplets would grow
and percolate across the system, giving rise to superconductivity.
In the absence of doping, this cannot occur, and the STO substrate
is insulating. However, these droplets can interact with the itinerant
electrons at the interface. Through the proximity effect, superconducting
droplets in the STO substrate can induce a gap in the itinerant electrons
at the interface. Thus seeded, the itinerant electrons at the interface
can develop long-ranged or quasi-long-ranged superconducting order.
However, these itinerant electrons must move in the magnetic background created
by the localized electrons. Naively, the magnetism should destroy the superconductivity.
However, the strong spin-orbit coupling of the interface electrons allows
these two competing phenomena to coexist peacefully~\cite{GorkovRashba}. Spin-orbit coupling
mixes $s$-wave and $p$-wave superconductivity, so that the $s$-wave
superconductivity which is present in the droplets in STO can induce
a mixture of $s$-wave and $p$-wave superconductivity at the interface.
This mixture can tolerate a magnetic moment, unlike pure $s$-wave
superconductivity. This is very similar to the situation in proposals
of topological insulators in contact with $s$-wave superconductors \cite{FuKane, FuKane09, Cook'11},
superconductor-semiconductor heterostructures \cite{Sau09, Alicea_PRB10}, and
spin-orbit-coupled quantum wires \cite{1DwiresLutchyn, 1DwiresOreg}.
Therefore, according to our theory, even though superconductivity is $s$-wave
in STO, interfacial superconductivity is unconventional, as a result of magnetism
and spin-orbit coupling.

An especially exciting consequence of the unconventional nature of the interfacial superconductivity is the possibility of realizing Majorana fermion physics.  Indeed, the fabrication of narrow
quasi-one dimensional conducting channels on an otherwise insulating LAO/STO interface
\cite{Cen08, Levy1, Levy2} by `writing' them with an atomic-force microscope (AFM) tip suggests a natural implementation for the proposals of Refs.~\onlinecite{1DwiresLutchyn, 1DwiresOreg}.  In the following sections we show that our model naturally generates a `helical' interfacial band structure that can be driven into a topological phase when in proximity to ordinary s-wave superconductivity. As shown by Kitaev~[\onlinecite{1DwiresKitaev}], 1D spinless superconductor with p-wave (or effectively p-wave) pairing supports Majorana zero-energy modes at the ends. Although our droplet model can at best lead to quasi-long ranged superconducting order, signatures of Majorana physics still remain, as shown in Refs.~[\onlinecite{QMajorana, Sau1D, Meng1D}].

In Section \ref{sec:model}, we set up and justify
a model of spin-orbit-coupled itinerant electrons interacting with
Kondo spins. In Section \ref{sec:mean-field}, we give a saddle-point analysis
of a large-$N$ limit of this model.
In Section \ref{sec:dmrg}, we solve this model numerically by the density-matrix renormalization
group (DMRG) in the one-dimensional limit. In Section \ref{sec:superconductivity},
we analyze the superconducting proximity effect in the presence of a helical wire due to the presence of ferromagnetic Kondo interactions
and spin-orbit coupling. We show that such an interfacial superconducting state might support Majorana zero-energy modes. We conclude in Sec.~\ref{sec:discussion} with the discussion of our results and proposing a schematic phase diagram for LAO/STO interface.

\section{Spin-Orbit-Coupled Ferromagnetic Kondo Model}
\label{sec:model}

We will now argue for a model of spin-orbit coupled electrons interacting with localized spins to describe the LAO/STO interface.  The emerging consensus is that the electrons active at the LAO/STO interface come from the $t_{2g}$ bands of Ti $3d$ orbitals in the STO \cite{Ariando11,Joshua11,Kopp10,Popovic,Pavlenko12,Delugas11, Salluzzo09, Pentcheva06, Mannhart08}.  Furthermore, band structure calculations \cite{Joshua11} and density functional theory \cite{Delugas11} suggest a picture of successive Ti $3d$ sub-bands near the interface being occupied as LAO thickness is increased (or gate voltage decreased).

Due to the inversion symmetry breaking at the interface, the lowest sub-band is predicted to be
$d_{xy}$.  It contains most of the charge required by the polarization catastrophe, but these electrons are thought to be localized, as seen from the low mobile carrier densities extracted from Hall transport measurements.  We thus model these electrons as localized spins.  Further indirect evidence for a picture involving localized spins in LAO/STO comes from transport experiments on a related system consisting of pure STO, with the doping effect of the LAO simulated by an polarized gel overlayer \cite{Goldhaber-Gordon11}.  The longitudonal resistance in this system exhibits a Kondo minimum as a function of temperature, indicative of the presence of localized
impurity spins.

The lowest extended interface states are thought to have $d_{xz}$ and $d_{yz}$ symmetry.  Due to the geometry of their orbitals, their Fermi surfaces are highly anisotropic, with heavy and light carrier directions.  Spin-orbit coupling also plays an important role in determining the electronic band-structure, with the authors of Ref. \onlinecite{Joshua11} arguing for an
atomic spin orbit (ASO) effect of about $10 {\rm meV}$.
Furthermore, there is a Rashba contribution arising from the broken inversion symmetry of the interface and gating.  Its magnitude is expected to be dependent on the gate voltage and the details of the sample, but in Ref. \onlinecite{Caviglia10} a value of
$\alpha \approx 10-50\, {\rm meV} \AA$ is obtained through a fit to a weak anti-localization measurement.  (Larger values close to $50 {\rm meV} \AA$ are obtained for
larger gate voltages; smaller values, for smaller gate voltages.)
The Rashba nature of the coupling was deduced from the dependence
of the spin relaxation time on the elastic scattering time.
The electronic band-structure for the two dimensional interface
depends on the precise ratio of Rashba and ASO coupling.

Motivated by recent experiments fabricating one dimensional conducting channels on otherwise insulating LAO/STO interfaces \cite{Levy1, Levy2}, as well as by our desire to realize Majorana physics, we find it useful to examine one-dimensional channels at the LAO/STO interface.
In a 1D channel, the anisotropy of the $d_{xz}, d_{yz}$ orbitals suggest that for a very narrow channel along the $x$ direction, transport should primarily be through the $d_{xz}$ states \cite{Triscone12}. Therefore, the neglect of the $d_{yz}$ band is justified in this situation.
In Ref. \onlinecite{Levy1,Levy2}, conducting channels of thickness $\sim 10 \,{\rm nm}$
are constructed.  For an effective mass of order the electron mass, such channels correspond to a transverse confining energy of $\sim 2 \, {\rm meV}$, and hence in principle we need to consider many subbands at the energy scales in which we are interested.  However, if narrower conducting wires could be constructed (e.g. $2 \,{\rm nm}$, corresponding to $\sim 40 \, {\rm meV}$), a one dimensional model would be more readily applicable.

We are thus led to a analyze a minimal model of a single spinful band interacting
with a large density of localized spins. The localized spins, which give
the dominant contribution to the magnetic moment come from the $d_{xy}$ band.
However, their tendency to order is driven primarily by their interaction with $d_{xz}, d_{yz}$
electrons. Although the experiment of Ref. \onlinecite{Goldhaber-Gordon11}
suggests an antiferromagnetic coupling
between conduction and impurity spins, we believe that it is more natural,
according to Hund's rule, take the interaction between $d_{xy}$ electrons and
$d_{xz}, d_{yz}$ electrons to be ferromagnetic.
In either case, the Hamiltonian takes the form
\begin{align} \label{Hamiltonian}
H&= \sum_{k_x, \lambda, \lambda'} \left[ \varepsilon(k_x)\delta_{\lambda,\lambda'} c_{k_x, \lambda}^\dag c_{k_x, \lambda} + \alpha k_x \sigma_{\lambda, \lambda'}^y c_{k_x, \lambda}^\dag c_{k_x,\lambda'}\right] \nonumber \\ &+ J \sum_{k_x, q, \lambda, \lambda'} c_{k_x+q,\lambda}^\dag \vec{\sigma}_{\lambda, \lambda'} c_{k_x, \lambda'} \cdot \vec{S}(q)
\end{align}
where $\sigma_i$ is the Pauli matrix acting on a spin degree of freedom, $\vec{S}$ corresponds to classical spins due to localized magnetic impurities, and we have suppressed spin indices in the notation.  In a lattice version of this Hamiltonian the kinetic term is modeled by a nearest neighbor hopping $t$, whose magnitude is determined by band structure calculations \cite{Popovic, Pavlenko12} to be about $0.2 \, {\rm eV}$.  From magnetoconductance experiments \cite{Caviglia10, Triscone12} we take $\alpha \approx 10-50\, {\rm meV} \AA$, which translates to a lattice spin orbit coupling $\alpha a^{-1} \sim t/20$ or even as large as $t/4$ (see below).  The final term is the interaction between the itinerant electrons and localized spins.  Although we believe that it is more natural,
according to Hund's rule, to have a ferromagnetic coupling, $J<0$, we will also, for completeness,
consider the case of antiferromagnetic coupling, $J>0$, with a magnitude of roughly one third of the bandwidth, as obtained in Ref. \onlinecite{Goldhaber-Gordon11}.

\begin{figure}
\includegraphics[width = 9cm]{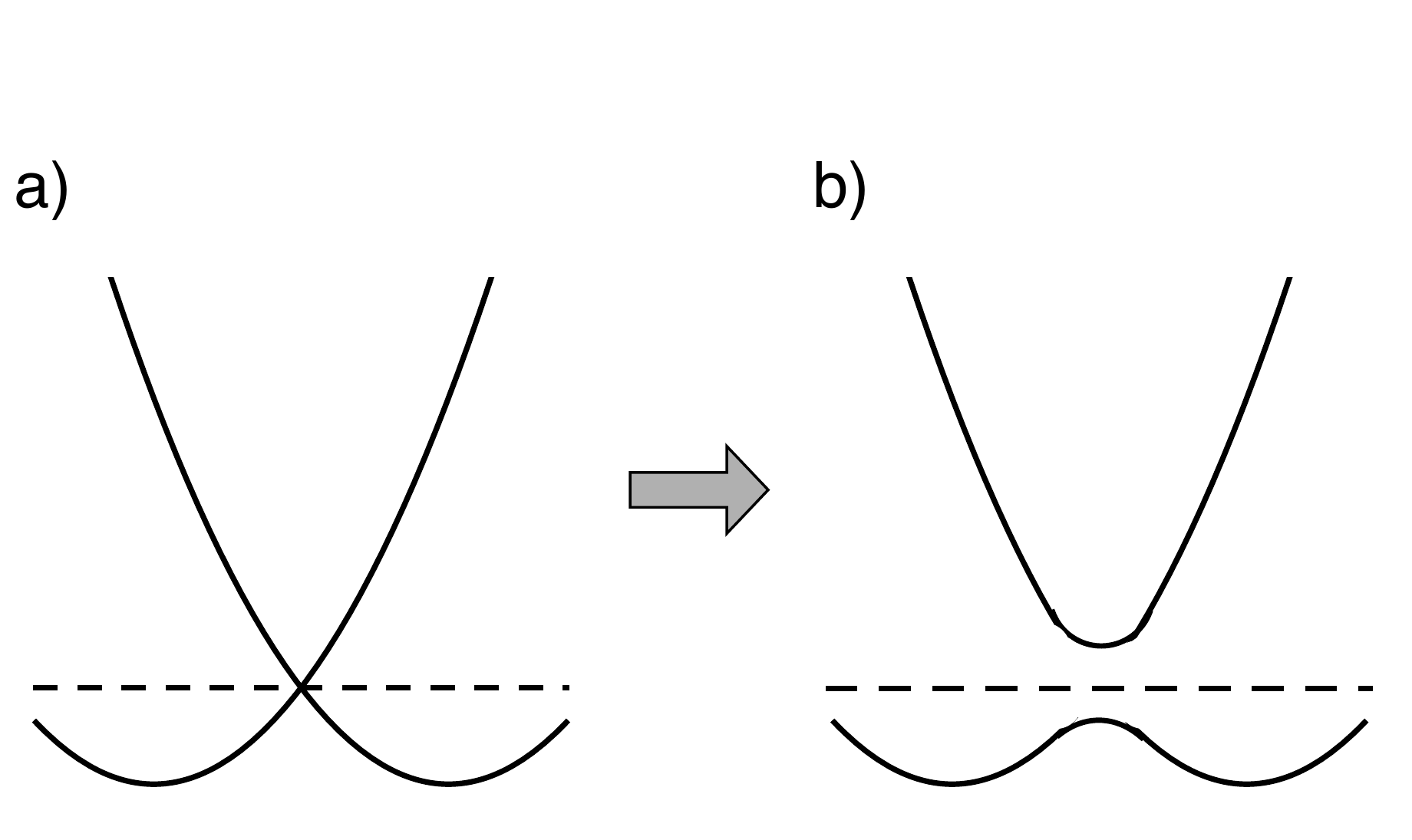}
\caption{(a) Sample one dimensional electronic band-structure.  We expect a magnetic instability at ordering wave-vector $q=0$, where a gap can open in the spectrum at the crossing
point between the two bands. When this occurs, the band structure takes the
form in (b)}
\label{1d_Rashba}
\end{figure}

Motivated by recent SQUID \cite{Bert11} and torque magnetometry \cite{Li11} experiments,
which find at least local ferromagnetism, we are interested in examining the magnetic instabilities of the Hamiltonian (\ref{Hamiltonian}).  The one dimensional spin-orbit coupled electronic band-structure in figure \ref{1d_Rashba} shows that it is natural to expect ordering at wave-vector $q=0$, since it is there that a gap can be opened and the electronic energy lowered. Of course,
ordering could also occur at the wavevectors that connect various pairs of Fermi points.
In the next section we will study a large-$N$ limit of this model and find a tendency toward in-plane ferromagnetism, consistent with [\onlinecite{Bert11, Li11}].  In the following section we will reach a similar conclusion with DMRG.

\section{Large-$N$ Analysis of the Model}
\label{sec:mean-field}

\begin{figure}
\includegraphics[width = 9cm]{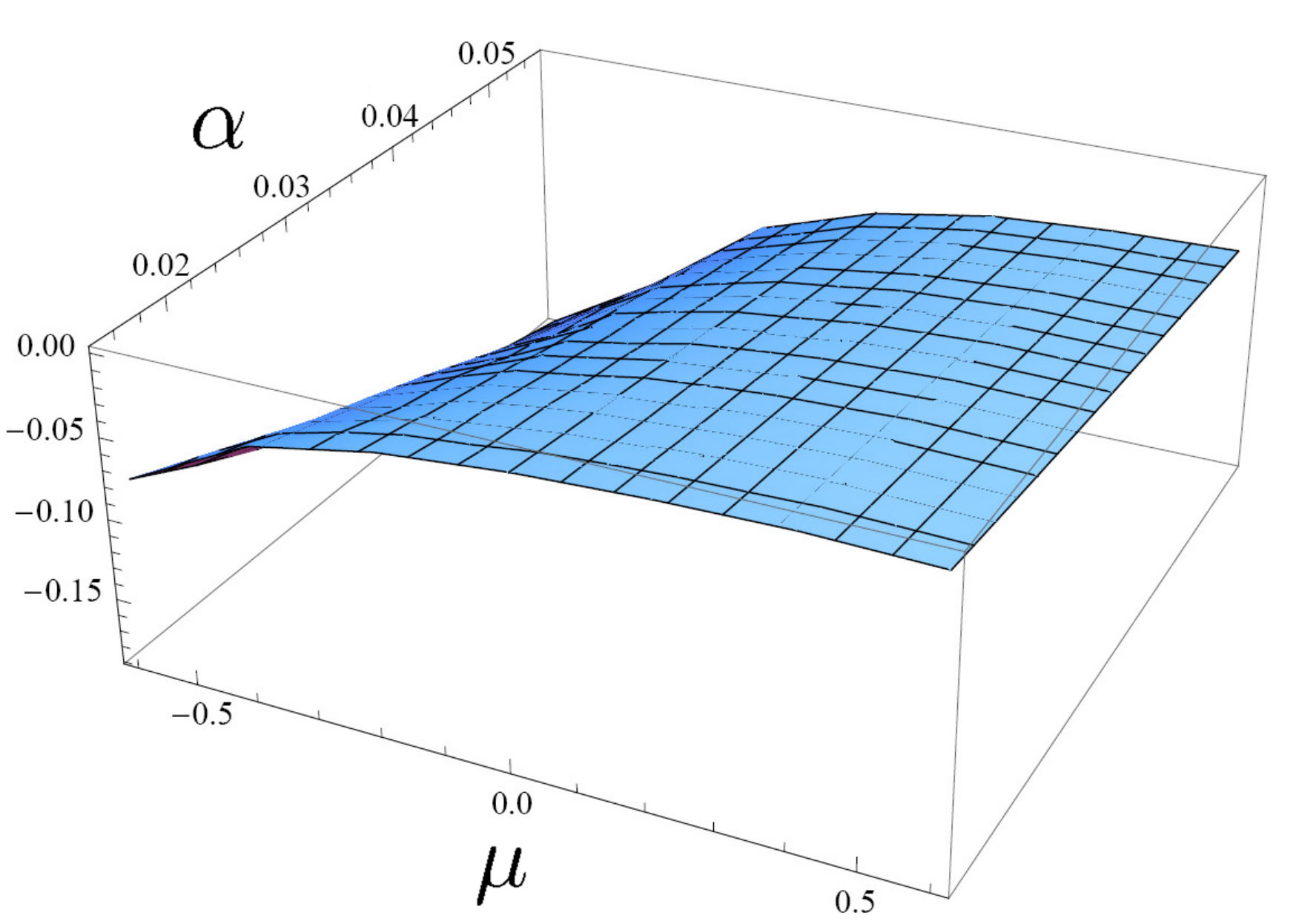}
\caption{$-(\delta E)/ |E|$, where $\delta E$ is defined in (\ref{deltaE}), plotted at $J = 0.4$ as a function of chemical potential $\mu$ and spin-orbit coupling $\alpha$ for a one dimensional conducting wire aligned in the $y$ direction sitting on the $2$ dimensional $xy$ plane.  We note that this quantity is everywhere negative, implying a propesity towards magnetic ordering in plane, perpendicular to the wire {\emph i.e.} along the $x$ direction.  Other values of $J$ give similar results.}
\label{1d_frac}
\end{figure}

We now analyze magnetic instabilities of a generalization
of the Hamiltonian (\ref{Hamiltonian}) to $N$ species of fermions
$c^a_{k_x}$, with $a=1,2, \ldots, N$:
\begin{align}
\label{Hamiltonian}
H &= \sum_{k_x, \lambda, \lambda',a} \left[ \varepsilon(k_x) c^{a\dag}_{k_x\lambda} c^a_{k_x, \lambda}\delta_{\lambda, \lambda'} + \alpha k_x  \sigma^y_{\lambda, \lambda'} c_{k_x, \lambda}^{a\dag} c^a_{k_x, \lambda'}  \right] \nonumber \\
& \hskip 1 cm + J \sum_{k_x,q, \lambda, \lambda' } c^{a\dag}_{k_x+q, \lambda} \vec{\sigma}_{\lambda, \lambda'} c^a_{k_x, \lambda'} \cdot \vec{S}_q
\nonumber \\
&= \sum_{k_x, \lambda, \lambda', a} c_{k_x\lambda}^{a\dag} M(k_x, \vec S )_{\lambda \lambda'} c^a_{k_x \lambda'}
\end{align}
where the matrix $M(k_x, \vec S)$ is given by
\begin{align}\label{eq:M_largeN}
M(k_x, \vec S)\!=\!\left[\begin{array}{cc} \varepsilon(k_x) \!+\! J S_z & i\alpha k_x + J (S_x + i  S_y) \\
                -i\alpha k_x + J (S_x - i S_y) & \varepsilon(k_x) - J S_z \end{array} \right].
\end{align}
We now integrate out fermions in Eq.\eqref{Hamiltonian}:
\begin{align}
\exp\left(-S_{\rm eff}[\vec{S}]\right) &= \int Dc\, dc^\dagger \, e^{-{S_0}[\vec S] - \int d\tau \sum_k
(c^{a\dag}_{k_x} i \partial_\tau c^a_{k_x} - H)}\cr
&= e^{-{S_0}[\vec{S}] -N \text{tr}\ln\left(i\omega_n - M(k_x, \vec S)\right)}
\end{align}
where ${S_0}[\vec{S}]$ is the Berry phase term for the localized spins.
In the large-$N$ limit, the functional integral
\begin{align}
Z[\vec{S}] = \int D S \, \exp\left(-{S_0}[\vec{S}] - N \text{tr}\ln \left[ i\omega_n - M(k_x, \vec S) \right]\right)
\end{align}
is equal to its saddle-point value. The saddle-point equations are given by:
\begin{align}
\frac{\partial}{\partial \vec{S}}\left({S_0}[\vec{S}] - N \text{tr}\ln \left[ i\omega_n - M(k_x, \vec S) \right]\right)= 0.
\end{align}
The ${S_0}[\vec S]$ term is $O(1)$ and is much smaller than the second term which is $O(N)$, so it can be neglected. Since the temperature of interest is much lower than the energy scales associated with the couplings in \eqref{Hamiltonian}, we can effectively set it to zero and convert Matsubara sum to an integral. Thus, we arrive at the following mean-field equations:
\begin{align}
\frac{\partial}{\partial \vec{S}} \int \frac{d\omega}{2\pi} \frac{dk_x}{2\pi} \log\left[\right(i\omega - E^+(k_x)\left)\right(i\omega - E^-(k_x, \vec S)\left)\right]=0,
\end{align}
where $E^{\pm}(k_x)$ are the two eigenvalues of $M(k_x, \vec S)$:
\begin{align} \label{epem}
E^{\pm}(k_x, \vec S) &= \varepsilon(k_x) \pm Q(k_x,\vec S) \\
Q(k_x,\vec S) &=\sqrt{J^2 S_z^2 + J^2 S_x^2+ (\alpha k_x + J S_y)^2}
\end{align}
The explicit evaluation of derivatives yields
\begin{align} \label{saddlep}
\!\int \frac{d\omega}{2\pi} \frac{dk_x}{2\pi} \frac{\partial Q(k_x, \vec S)}{\partial \vec{S}} \left[\frac{1}{i\omega \!-\! E^+(k_x, \vec S)} - \frac{1}{i\omega \!-\! E^-(k_x, \vec S)} \right]\!=\!0.
\end{align}
One can notice that we only obtain a non-zero contribution if $E^+(k_x,\vec S)$ and $E^-(k_x, \vec S)$ have opposing signs, so that the saddle point equation reduces to
\begin{align}
\int \frac{dk_x}{2\pi} \left[ \Theta(-E^+(k_x, \vec S)) - \Theta(-E^-(k_x, \vec S)) \right] \frac{\partial Q(k_x,\vec S)}{\partial {\vec{S}}} = 0.
\end{align}
Here $\Theta(x)$ is the unit step function ($\Theta(x) = 1$ for $x>0$) and
\begin{align}
\frac{\partial Q(k_x, \vec S)}{\partial {\vec{S}}} = \frac{J}{Q(k_x, \vec S)} \left( \begin{array}{c}  J S_x \\  \alpha k_x + J S_y \\ J S_z \end{array} \right)
\end{align}
The $\vec{S}$-dependent contribution to the energy of a particular background spin configuration at one loop reads
\be
E \propto \int \frac{dk}{2\pi} \left(E_k^+ \Theta(-E_k^+) + E_k^- \Theta(-E_k^-) \right)
\ee
To diagonalize, we take nearest neighbor hopping with amplitude normalized to $1$, resulting in a kinetic term
\be
\varepsilon (k_x)= \cos(k_x)-\mu
\ee
For convenience we also absorb ${\vec{S}}$ into $J$: $\vec{J} \equiv J \vec{S}$, and drop it from the following discussion.  We then compute the one loop energy as a function of $\vec{J}, \mu$, and $\alpha$, and determine the propensity for magnetic ordering in various directions.  From (\ref{epem}) we see that the $x$ and $z$ directions become equivalent at one loop, so it suffices to set $J_x=0$ and work with nonzero $J_y$ and $J_z$.  Symmetry considerations show that $\vec{J}=0$ is an extremum of the one loop energy.  In fact, we empirically see that it is a global maximum, and that the energy is unbounded from below, becoming more negative with increasing $|J|$.  This makes sense since the saddle-point Hamiltonian treats ${\vec{S}}$ as a classical field with no dynamics, and larger $|\vec{S}|$ leads to lower electronic energy.  Physically, we expect that $|\vec{S}|$ ultimately saturates.  For the purposes of the saddle-point approximation we pick an appropriate value for $|{\vec{J}}|$ and evaluate the energy difference between ordering in the $y$ and $z$ directions:
\begin{align} \label{deltaE}
 \delta E(\alpha, \mu) &= E(J,0,\alpha,\mu)-E(0,J,\alpha,\mu) \nonumber \\
&\approx J^2 \left(c_z(\alpha,\mu) - c_x(\alpha,\mu) \right)
\end{align}
In figure \ref{1d_frac} we plot $\delta E / |E|$ at a specific value of $J$, and note that it is everywhere negative.  Although the percentage difference in energies is small, it is robustly negative over a very large range of physical values of $\mu, \alpha$, and $J$.  We conclude that in saddle-point approximation the spins prefer to develop magnetic order in the (in-plane) $y$ direction.

We can perform a similar analysis for a two dimensional version of (\ref{Hamiltonian}), obtaining a similar result: the spins prefer to order in-plane in the large-$N$ limit.

One very interesting part of our analysis is that we find ferromagnetic order developing
at {\it weak coupling}. The underlying reason for this is that the band structure
in the presence of Rashba spin-orbit coupling has a crossing at $k=0$; a small
magnetic moment opens a gap there. This occurs even at arbitrarily weak coupling,
if the chemical potential passes through this crossing. This is very similar to
the case of other Fermi surface instabilities, such as the BCS instability
or density-wave ordering for nested Fermi surfaces. If the chemical potential does
not pass through the $k=0$ crossing, then a small minimum coupling must be exceeded,
as in the case of small detuning away from a nested Fermi surface. This scenario
stands in stark contrast to the usual case of the Stoner instability: ordinarily, ferromagnetism
does not open a gap at the Fermi surface, and only occurs when the coupling exceeds
the inverse of the density of states.

\begin{figure}[tbp]
\centerline{
    \includegraphics[width=3.6in]{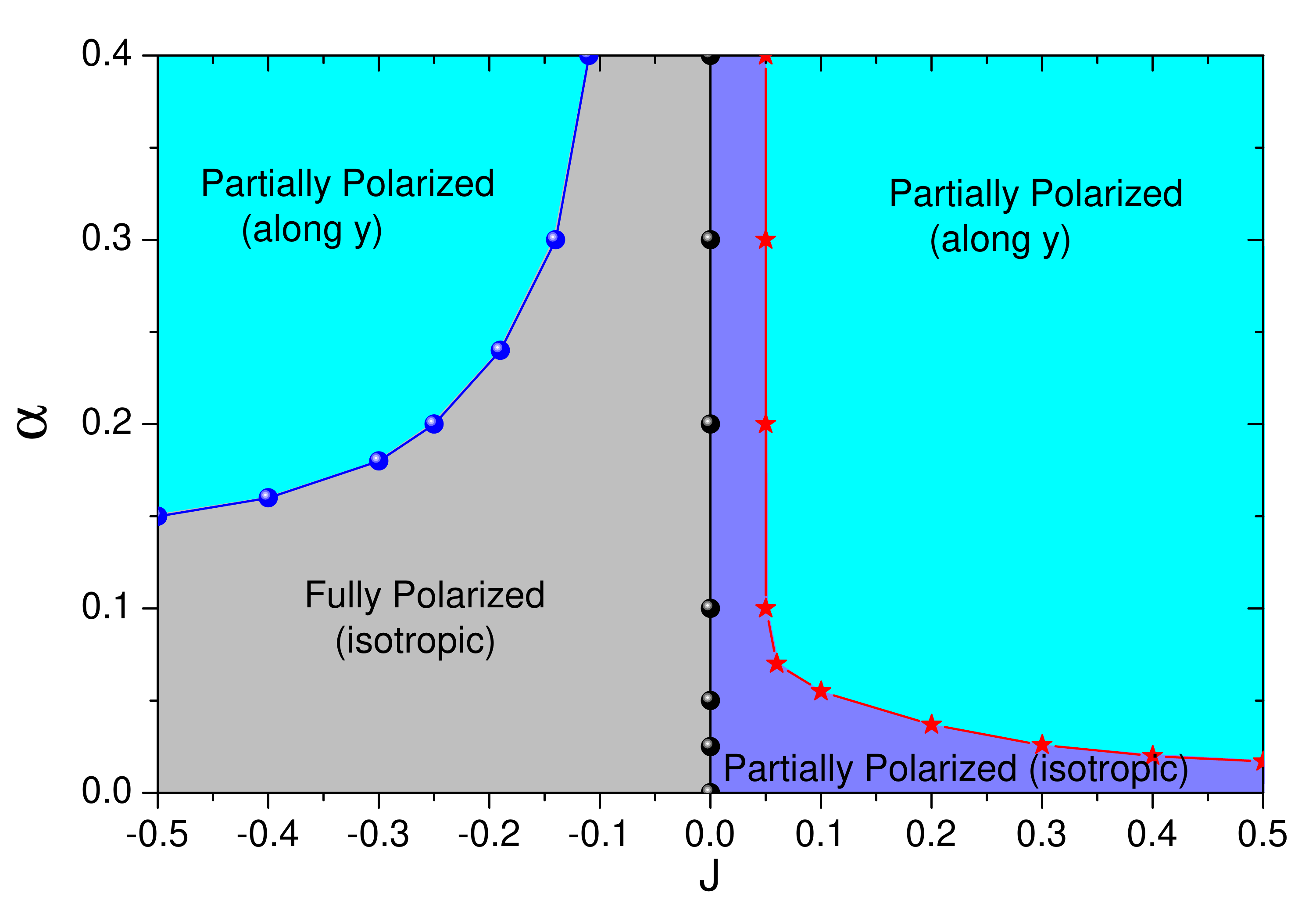}}
\caption{(color online) Ground state phase diagram of the model
Hamiltonian in Eq.(\ref{Eq:HamiltonianDMRG1D}) at filling
$\rho=1/6$, determined by accurate DMRG simulations with system size
up to $N=192$ sites. Changing coupling parameters $J$ and
$\alpha a^{-1}$, three different phases are found, including the fully
polarized phase, the partially polarized phase, as well as the
easy-plane partially polarized phase. Here $J_{\rm chain}=0.1t$ and
$U=40t$.} \label{Fig:PhaseDiagramDMRG1D}
\end{figure}

\section{DMRG Solution of the 1D Limit}
\label{sec:dmrg}%

We now consider a 1D lattice Hamiltonian describing itinerant electrons coupled to localized impurity spins:
\begin{eqnarray}
H &=& -t{\sum_{i, \alpha}}(c^+_{i\alpha}c_{i+1\alpha} + h.c.) +
J {\sum_{i, \alpha, \beta}}\vec{S}_i\cdot
c^+_{i\alpha}\vec{\sigma}_{\alpha\beta}c_{i\beta} \nonumber\\ %
&-&{J_{\rm chain}}{\sum_i}\vec{S}_i\cdot \vec{S}_{i+1}\label{Eq:HamiltonianDMRG1D}\\
&+& \frac{\alpha}{a}{\sum_i}
(c^+_{i\uparrow}c_{i+1\downarrow} - c^+_{i\downarrow}c_{i+1\uparrow}+ h.c.)\nonumber\\%
&+& U{\sum_i} n_{i\uparrow}n_{i\downarrow} -h_z{\sum_i} (S^z_i +
\tau^z_i) -h_y{\sum_i} (S^y_i + \tau^y_i). \nonumber
\end{eqnarray}
Here $c^+_{i\alpha}$ ($c_{i\alpha}$) is the electron creation
(annihilation) operator with spin index
$\alpha=(\uparrow,\downarrow)$ at site $i$; $\vec{S}_i$ is the
$S=\frac{1}{2}$ spin operator, representing the localized magnetic
moment; $t$ denotes the nearest neighbor (NN) tunneling matrix element (henceforth we set $t=1$) and $a$ is the lattice constant. $J$ is the Kondo coupling between
localized magnetic moments and itinerant electrons, $J_{\rm chain}$ is the
ferromagnetic exchange coupling between NN localized magnetic
moments, and $\alpha$ is the spin-orbit coupling. $U$ is the on-site Hubbard
repulsion for the itinerant electrons, and we have also included Zeeman terms for both the localized spin and itinerant electron spin along the $z$ and $y$ directions. In the Hamiltonian~\eqref{Eq:HamiltonianDMRG1D}, impurity spins have their own dynamics and the ground-state of the is determined by taking into account both electron and spin degrees of freedom on equal footing. If we use mean-field approximation for impurity spins, i.e. $\langle \vec S_i \rangle=\vec{S}(x_i)$, and neglect electron-electron interaction, we obtain the Hamiltonian \eqref{Hamiltonian} considered in the previous section. As we show below, our conclusions regarding preferred magnetization persist in the strongly interacting limit $U \gg t$.

\begin{figure}[tbp]
\centerline{
    \includegraphics[width=3.4in]{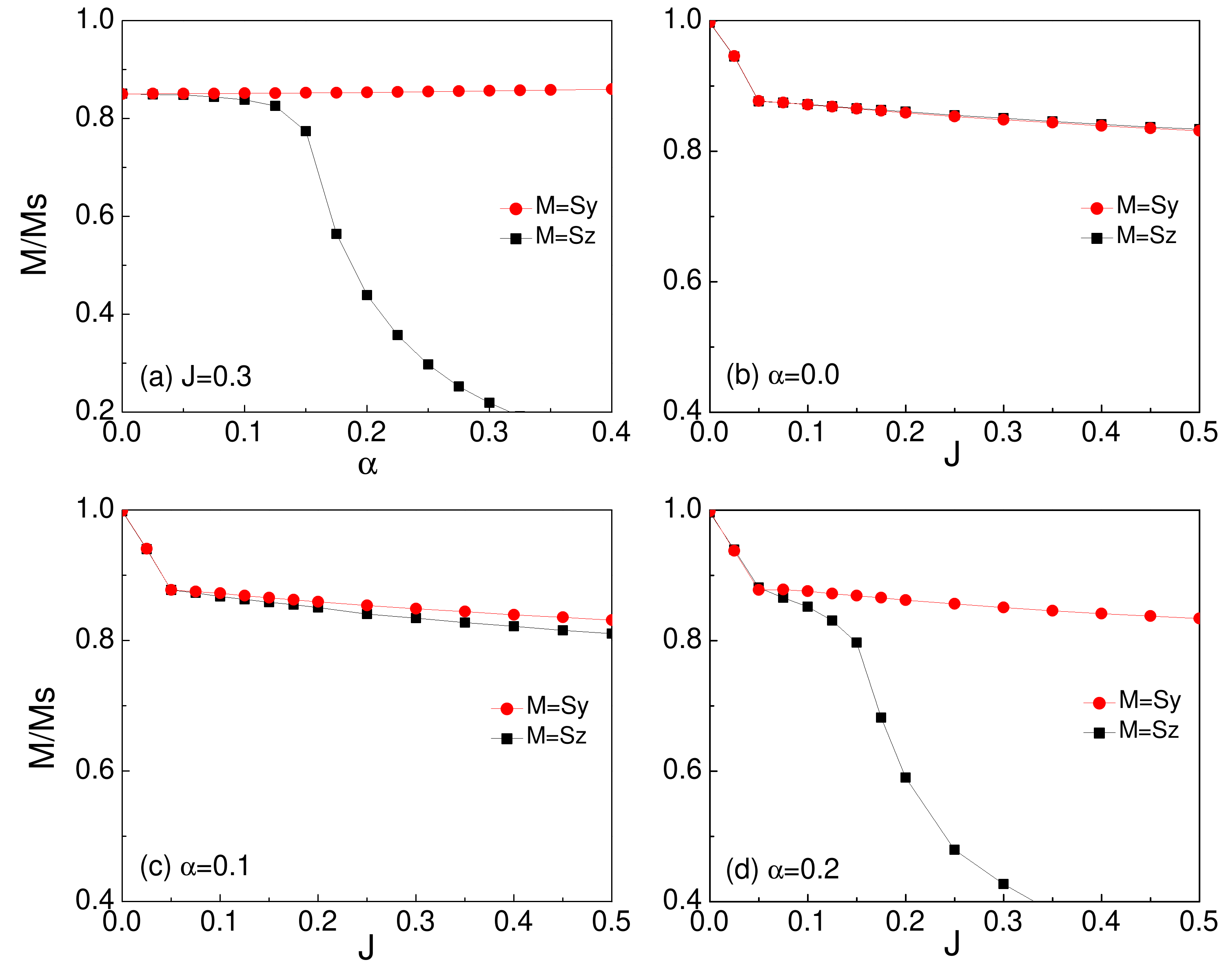}
    }
\caption{(color online) Relative polarization $M/M_s$ as functions
of $\alpha a^{-1}$ and $J$, along $y$ direction ($M=S^y$) and $z$
direction ($M=S^z$), for the system in
Eq.(\ref{Eq:HamiltonianDMRG1D}) at filling $\rho=1/6$ and system
size $N=192$ sites. Relative polarization $M/M_s$ for $J=0.3$ in
(a), $\alpha a^{-1}=0.0$ in (b), $\alpha a^{-1}=0.1$ in (c), and $\alpha a^{-1}=0.2$ in
(d). Here $J_{\rm chain}=0.1t$, $U=40t$, and $M_s$ is the saturated
magnetization. Note that the polarization is induced by applying a
small magnetic field $h_y=0.005t$ along $y$ direction or
$h_z=0.005t$ along $z$ direction separately.}
\label{Fig:MagnetizationYZ}
\end{figure}

In our picture, most of the electrons required by the polarization
catastrophe argument become localized $d_{xy}$ spins, with a much lower density of itinerant electrons.  Therefore, we focus on the low density case, taking $\rho = 1/6$ for the sake of concreteness in most of our calculations.  The coupling parameters are taken to have
values $U=40t$ and $J_{\rm chain}=t/10$. (However, our results are not
very sensitive to the value of $J_{\rm chain}$.) We then map out the phase diagram as
a function of $\alpha a^{-1}$ and $J$. By the arguments given in Sec. \ref{sec:model},
we expect that $\alpha a^{-1} \approx t/40$ and $J\approx -0.3t$.
However, given that there is some uncertainty in these parameters, it behooves us at this
stage to see how the physics of our Hamiltonian changes as we vary them.
We have included small magnetic fields in the $y$- and $z-$directions.
These fields are necessary to break time-reversal symmetry and rotational
symmetry about the $S^y$-axis; otherwise, we would necessarily find
$\left\langle S^y \right\rangle = \left\langle S^z \right\rangle = 0$. In an infinite system,
but in the absence of these symmetry-breaking fields, the system can spontaneously choose
to order along $S^y$ or $-S^y$ or it could spontaneously pick a direction
in the ${S^z}-{S^x}$ plane (if it orders at all).
Employing the unbiased density matrix renormalization group\cite{White1992} method, we determine the ground-state phase diagram of the system
(\ref{Eq:HamiltonianDMRG1D}).

We find the following phase diagram, depicted in Fig.\ref{Fig:PhaseDiagramDMRG1D}.
For $J$ large and negative (i.e. a ferromagnetic coupling between localized and itinerant
spins) and $\alpha a^{-1}$ large, the system is partially-polarized
and the moment points in the $y$-direction (i.e. in plane, but perpendicular to the nanowire, and corresponding to the light blue region in the upper left
corner of Fig.\ref{Fig:PhaseDiagramDMRG1D}). For $J<0$ but lying below the blue
phase boundary in Fig. \ref{Fig:PhaseDiagramDMRG1D}, we have a fully-polarized
phase in which the spins can point equally-well in any direction (the grey region near the middle
of Fig.\ref{Fig:PhaseDiagramDMRG1D}). For $J>0$, there is a partially-polarized phase
at small $J$ or small $\alpha a^{-1}$ (narrow purple region). In this region, the spins
can point can point equally-well in any direction. Finally, if $J$ is positive and either
$J$ or $\alpha a^{-1}$ is large, then the system is partially-polarized and the spins
point in the $y$-direction (light blue region in upper right of Fig. \ref{Fig:PhaseDiagramDMRG1D}).

\begin{figure}[tbp]
\centerline{
    \includegraphics[height=2.9in,width=3.4in]{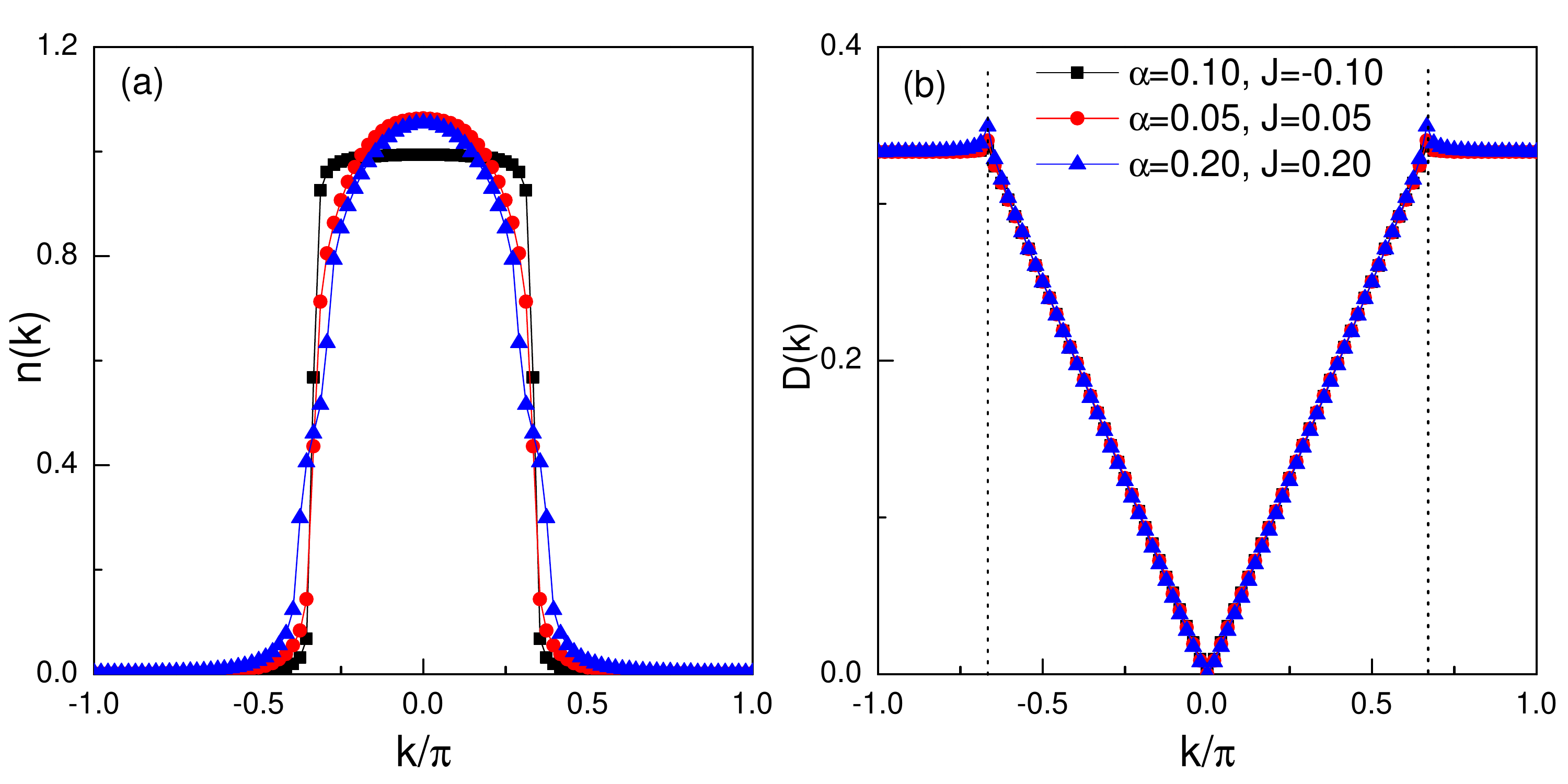}}
\caption{(color online) The Fermi surface of our 1D model at three points
in the phase diagram. (a) The occupation number $n(k)=
\langle c^\dagger_\uparrow c^{}_\uparrow +
c^\dagger_\downarrow c^{}_\downarrow\rangle$, which shows the region
of momentum space occupied by the filled Fermi sea. (b) The equal-time
density-density correlation function, which has singularities at $2k_F$.}
\label{Fig:Fermi-surface}
\end{figure}

We now examine the magnetically-ordered state in more detail.
The occupation number $n(k)=
\langle c^\dagger_\uparrow c^{}_\uparrow +
c^\dagger_\downarrow c^{}_\downarrow\rangle$ clearly shows a filled
Fermi sea with approximately one electron in each occupied state,
with $|k|<k_F$, as illustrated in Fig. \ref{Fig:Fermi-surface}a,
at the parameter values listed in Fig. \ref{Fig:Fermi-surface}b.
The Fermi wavevector is consistent with $k_F =\pi/3$,
in agreement with Luttinger's theorem.
The structure factor $\rho(q) \rho(-q)$ shown in
in Fig. \ref{Fig:Fermi-surface}b has cusps at $\pm 2k_F$,
from which we can more precisely extract the Fermi wavevector $k_F =\pi/3$.
The occupation numbers and Luttinger volume are consistent with
the chemical potential depicted in Figure \ref{1d_Rashba}.
There are two Fermi points and there is a single state at each Fermi
point because the spin is locked to the momentum.
Such a 1D electron gas is often called a ``helical wire''.
(Note that our system is not simply fully spin-polarized;
we have checked that, at these parameter values,
neither $\langle c^\dagger_\uparrow c^{}_\uparrow \rangle$ nor
$\langle c^\dagger_\downarrow c^{}_\downarrow \rangle$ is ever equal to
one, so the spins are not polarized in the $z$-direction;
they are also not equal to each other, so the spins are not
polarized in the $x$- or $y$- directions.)
Therefore, an odd number of bands (in fact, just one) crosses the Fermi surface;
as we discuss in Section \ref{sec:superconductivity}, this means that the
system is primed for the development of topological superconductivity.

\section{Proximity-Induced Superconductivity}
\label{sec:superconductivity}

We propose the following picture for superconductivity at the LAO/STO
interface. Our starting point is superconductivity in STO, which occurs
when insulating STO is doped. We assume that insulating STO has small islands
or droplets of local superconductivity, which are too far apart and too weakly-coupled
to develop long-ranged superconducting order. We suppose that these droplets are caused
by unintentional local defects in STO. When STO is doped, the islands
of local superconductivity grow in size and become more strongly
coupled, until long-ranged superconducting order sets in. However,
the presence of an interface with LAO changes matters. Itinerant electrons
at the interface can mediate a coupling between superconducting droplets in STO
that are close to the interface. As we will show, this can enable superconductivity to
develop even when the droplets are too weakly-coupled to percolate across
STO on their own.

We show how this can occur with a calculation in a simplified model.
We suppose that there are some superconducting droplets
in STO that are near the LAO/STO interface. In each droplet,
a single-particle gap is assumed to be well-developed, but the interactions
between the droplets are assumed to be too weak for superconducting order
to set in. A 1D channel at the interface
couples to a subset of these droplets, which form a linear array.
The 1D channel induces interactions between the droplets, so that
the linear array can be modeled as a 1D spin-gapped electron
system, which we assume to be just slightly on the disordered side of the
Kosterlitz-Thouless transition. (Since doped STO superconducts, this
is a reasonable approximation.)
The coupling between this system and a 1D channel at the LAO/STO
interface can nudge the system into the basin of attraction of the quasi-long-range
ordered superconducting phase on the other side of the Kosterlitz-Thouless transition.
We thereby see how proximity to a metallic interface can stabilize long-ranged or
quasi-long-ranged superconducting order.

Once superconductivity is established in STO, it is induced
at the interface by the proximity effect.  This mechanism of establishing superconducting order works for generic conducting 1D channels.  However, the interesting scenario for us occurs when, due to Rashba spin-orbit coupling and ferromagnetism, the 1D channel realizes the helical band structure in Fig \ref{1d_Rashba}, and the induced superconductivity is topological.  Hence, we begin with a bosonized helical 1D channel at the interface, coupled to an array of isolated, {\it i.e.} effectively zero dimensional, droplets:

\begin{multline}
S = \frac{v_F}{2\pi} K_{w} \int dx \,d\tau\,
 \left[(\partial_x \theta)^2+{v_F}^{-1}(\partial_\tau \theta)^2\right]\\
 + \frac{1}{2U_j}{\sum_j} (\partial_\tau {\theta_j})^2
 + {\Delta_{P,j}}{\sum_j} \left(e^{2i\theta({x_j})}\,e^{-i\sqrt{2}\theta_j} + \text{c.c.}\right)
\end{multline}
Here, $v_F$ is the Fermi velocity in the 1D channel and $K_w$ is its
Luttinger parameter. The 1D channel is assumed to have repulsive interactions,
so ${K_w}<1$. The factor of $\sqrt{2}$ in the exponent in the third line
has been inserted for later convenience so that it agrees with the convention
for the charge boson of a spin-gapped electron system
(see, for instance, Ref. \onlinecite{QMajorana}).
The droplets are assumed to have an average spacing $a$,
an average Josephson coupling ${\Delta_P}$ to the 1D channel, and an
average charging energy $U$. We will
neglect random variations in the spacing between droplets, in the Josephson
couplings, and in the charging energies, and simply set ${x_j}=ja$,
${\Delta_{P,j}}={\Delta_P}$, and ${U_j}=U$.
Random variations in these parameters are certainly important in the physical system
but are an unnecessary complication for this calculation.
Note that if ${\Delta_P}=0$, then the droplets are completely decoupled from
each other and there is no quasi-long-ranged superconducting order.

We now integrate out fluctuations of $\theta$ at length scales shorter than
$\ell$. This generates a coupling between droplets. At length scales much longer
than $\ell$, we can take the continuum limit for the array of droplets, thereby
leading to the following effective action:
\begin{multline}
S = \frac{1}{2\pi}\int dx \,d\tau\,\biggl(
 {v_F} K_{w} \left[(\partial_x \theta)^2+{v_F}^{-2}(\partial_\tau \theta)^2\right]\\
 + {v} K_{\rho}
 \left[(\partial_x \theta_{\rho})^2+v^{-2}(\partial_\tau \theta_{\rho})^2\right]\\
 - \frac{\Delta_P}{a} \cos\left(2\theta-\sqrt{2}\theta_\rho \right)\biggr)
\end{multline}
The droplets are now effectively described by a 1D wire with a spin gap;
$\theta_\rho$ is the charge boson for such a wire. Since we assume that there
is, initially, a weak interaction between the droplets and the 1D channel,
we assume that $K_\rho<1$. In other words,
although the wire has a spin gap, we do not assume that it can superconduct without
further mediation on the part of the 1D channel at the interface. Indeed, the parameters $K_{\rho}$ and $v$ can be related to the effective superfluid stiffness $\rho_s$ and compressibility $\kappa$ of the array: $K_{\rho}=2\pi \sqrt{A_w\rho_s \kappa}$ and $v=\sqrt{A_w\rho_s/\kappa}$ with being $A_w$ the cross-sectional area. We assume here that the superfluid stiffness is such that $K_{\rho}<1$.

If the two velocities were equal, $v=v_F$, one could analyze the model by forming the combinations
${\theta_\pm} = (\sqrt{2}\theta_\rho \pm 2\theta)/2$.
In terms of new variables $\theta_{\pm}$ the action reads
\begin{multline}
\label{fullaction}
S =  \frac{v}{2\pi}\int dx \,d\tau\,\biggl(
 \left(\mbox{$\frac{K_w}{4} + \frac{K_\rho}{2}$}\right)
 \left[(\partial_\mu {\theta_+})^2+(\partial_\mu {\theta_-})^2\right]\\
 - 2\left(\mbox{$\frac{K_w}{4} - \frac{K_\rho}{2}$}\right)
 \partial_\mu {\theta_+}\partial_\mu {\theta_-}
 - 2y \cos\left(2\theta_-\right)\biggr).
\end{multline}
Here, we have rescaled the time coordinate by $v$ and have introduced the
dimensionless parameter $y={\Delta_P}a/2v$.
We can now see, at a heuristic level, how the coupling between
the droplets and the 1D channel can stabilize quasi-long-ranged
superconducting order. Let us suppose, for a moment, that the coupling
$y$ is relevant. Then $\theta_-$ is pinned, and we can ignore its fluctuations.
Then we are left with $\theta_+$, which exhibits algebraically-decaying
superconducting order. This order is stable if
weak impurity-backscattering or, equivalently, vortex tunneling is irrelevant.
Because the wire is helical, we can only tunnel $\frac{hc}{e}$ vortices \cite{QMajorana}, i.e.
$\theta \rightarrow \theta +2\pi$. Since $\theta_-$ is pinned, this means that
$\theta_\rho$ must also wind $\theta_\rho \rightarrow \theta_\rho + 2\pi\sqrt{2}$.
Consequently, in such a process,
$\theta_+ \rightarrow \theta_+ +4\pi$.  The operator that accomplishes this is
$\sin (4{\phi_+})$, where $[{\phi_+}(x), \partial_x {\theta_+}(y)] = i \pi \delta(x-y)$.
This operator is irrelevant if $2{K_\rho} + {K_w} > 1$, and when this inequality
is satisfied, the system exhibits quasi-long-ranged superconducting order.
Note that this can be satisfied even if ${K_\rho}<1$
and ${K_w}<1$. So two systems, neither of which could sustain superconductivity
on their own, can develop superconductivity when in proximity to each other.
The key to this is the topological nature of the superconductivity: since
only $\frac{hc}{e}$ vortices can tunnel through a helical wire, the stability
condition is less strict than for an ordinary superconductor~\cite{Giamarchi_book}.

To support the aforementioned scenario, we need to show that
that Cooper-pair tunneling term is relevant. The corresponding RG equation
for $y$ is:
\begin{equation}
\label{eqn:y-flow1}
\frac{dy}{d\ell} = \left[2 - (K')^{-1}\right] y,
\end{equation}
where $(K')^{-1} \equiv \frac{1}{2K\rho} + \frac{1}{K_w}$.
If $(K')^{-1} < 2$, $y$ will grow from the initial small value $y(0)\ll 1$ to $y(l)\sim 1$ at which point $\theta_-$ gets pinned.
Given that $K'$ also flows under RG, we need to compute its flow equation and complete the system of RG equations for this model.
To do that we rewrite Eq. (\ref{fullaction}) in the following form:
\begin{multline}
\label{fullaction2}
S =  \frac{v}{2\pi}\int dx \,d\tau\,\biggl(
 {K_+}(\partial_\mu {\theta_+})^2+ {K_-}(\partial_\mu {\theta_-})^2\\
 - 2K_{+-} (\partial_\mu {\theta_+}\partial_\mu {\theta_-})
 - 2y \cos\left(2\theta_-\right)\biggr)
\end{multline}
where, initially, $K_{+} = K_{-} = \frac{K_w}{4} + \frac{K_\rho}{2}$,
$K_{+-}=\frac{K_w}{4} - \frac{K_\rho}{2}$.
The reason that we have introduced three couplings $K_{+}, K_{-}, K_{+-}$
when there are, seemingly, only two couplings $K_w$ and $K_\rho$
is that the RG flow for this theory will carry the system away from the initial point
$K_{+} = K_{-}$. A real-space RG calculation yields the following equations for $K_{-}, K_{+}, K_{+-}$:
\begin{equation}
\label{RGeq}
\frac{dK_{-}}{d\ell} = y^2 \, , \hskip 0.5 cm
\frac{dK_{+}}{d\ell} = \frac{dK_{+-}}{d\ell} = 0.
\end{equation}
The coupling $K'$ can be expressed in terms of $K_-$, $K_{+}$ and $K_{+-}$ (or equivalently in terms of $K_-$ and the initial parameters $K_{\rho}$ and $K_w$ since $K_{+-}$ and $K_+$ do not flow):
\begin{align}
K'&= K_{-} - K^2_{+-}/K_{+}=K_--\frac{(2K_{\rho}-K_w)^2}{4(2K_{\rho}+K_w)}\label{eq:Kprimeeq_a}.
\end{align}
Since $K'$ monotonically depends $K_{-}$,
the growth of $K_{-}$ under the RG flow results in
\begin{align}
\frac{dK'}{d\ell} & = y^2.\label{eq:Kprimeeq_b}
\end{align}
Thus, the $y$-coupling becomes
more and more relevant, and eventually pins $\theta_-$ as assumed above.

We now consider more general case of unequal velocities $v\neq v_F$. Proceeding as before,
we find that Eq. (\ref{fullaction}) can be written as
\begin{multline}
\label{fullaction3}
S = \frac{1}{2\pi}\int dx \,d\tau\,\Bigl({\tilde K}\left( {\tilde v} (\partial_x {\theta_-})^2
 +  {\tilde v}^{-1}(\partial_\tau {\theta_-})^2 \right)\\
+ {\tilde K}_{+}\left( {\tilde v}_{+} (\partial_x {\theta_+})^2
 +  {\tilde v}_{+}^{-1} (\partial_\tau {\theta_+})^2 \right) \\
 -2{\tilde K}_{+-}\left( {\tilde v}_{+-} (\partial_x {\theta_+})(\partial_x {\theta_-})
 +  {\tilde v}_{+-}^{-1} (\partial_\tau {\theta_+})  (\partial_\tau {\theta_-})\right)\\
  - 2y \cos\left(2\theta_-\right)
\Bigr)
\end{multline}
where the coupling constants are defined as
\begin{align}\label{eq:couplings}
{\tilde K_+} &= \frac{1}{4\sqrt{v v_F}}\sqrt{\left(\mbox{$ K_w{v_F} + 2K_\rho v$}\right)
\left(\mbox{$K_w v + 2K_\rho v_F$}\right)}\\
{\tilde v_+} &= \sqrt{v v_F\frac{2K_\rho v + {K_w}{v_F}}{2K_\rho v_F + {K_w}v}}\\
{\tilde K}_{+-} &= \frac{1}{4\sqrt{v v_F}}\sqrt{\left(\mbox{$K_w{v_F} - 2K_\rho v$}\right)
\left(\mbox{$ K_w v - 2K_\rho v_F$}\right)}\\
{\tilde v}_{+-}&=\sqrt{v v_F\frac{K_w{v_F} - 2K_\rho v}{K_w v - 2K_\rho v_F}}
\end{align}
The parameters $\tilde K$ and $\tilde v$ are initially equal to $\tilde K_+$ and $\tilde v_+$, respectively, but they flow under RG as explained above.

We now sketch the real-space RG procedure. We start out by integrating out short-distance modes, but allowing arbitrarily short times.
Thus, we have an effective action in which there are
modes $\theta(k,\omega)$ with $|k| < a^{-1}$ and
$|\omega|<\infty$, where $a^{-1}$ is the momentum cutoff. Following a real-space RG approach~\cite{Giamarchi_book}, we integrate shells $a<r<a s$ while keeping time integrals unconstrained and eventually rescale $r\rightarrow s r$, $\tau\rightarrow s \tau$. Here $s=e^{dl}$. Our RG procedure involves calculating the correlation function $\left\langle e^{2i{\theta_-}(x_1,\tau_1)} \cdot e^{-2i{\theta_-}(x_2,\tau_2)}\right\rangle$. To do it safely one has to normal order the exponent:
\begin{align}
\label{eqn:OPE}
e^{2i{\theta_-}(1)} \cdot e^{-2i{\theta_-}(2)}\,=:e^{2i[{\theta_-}(1)-{\theta_-}(2)]}: e^{-2\left\langle\left[\theta_-(1)-{\theta_-}(2)\right]^2\right\rangle}
\end{align}
where $\theta(1)\equiv \theta(x_1,\tau_1)$, and the average $\langle...\rangle$ is computed with respect to the bare action ($y=0$) defined in Eq.~\eqref{fullaction3}. Initially, when $\tilde K=\tilde K_+$ and $\tilde v=\tilde v_+$, the correlation function in Eq.~\ref{eqn:OPE} can be easily calculated
\begin{align}\label{eq:corr_fun}
e^{-2 \left \langle \left[\theta_-(x,\tau)-{\theta_-}(0,0)\right]^2\right \rangle }=\frac{a^{\frac{1}{2K_\rho}+\frac{1}{K_w}}}{({x^2}+{v^2}{\tau^2})^{\frac{1}{2K_\rho}} ({x^2}+{v_F^2}{\tau^2})^{\frac{1}{K_w}}}.
\end{align}
However, in general it is a complicated function of the coupling constants~\eqref{eq:couplings} as well as $\tilde K$ and $\tilde v$ . One can show that at the tree-level the RG equation for $y$ becomes
\begin{equation}
\label{eqn:y-flow2}
\frac{dy}{d\ell} = \left[2 - (K')^{-1}\right] y
\end{equation}
where, again, $(K')^{-1} \equiv \frac{1}{2K\rho} + \frac{1}{K_w}$.

\begin{figure}
\includegraphics[width = 9cm]{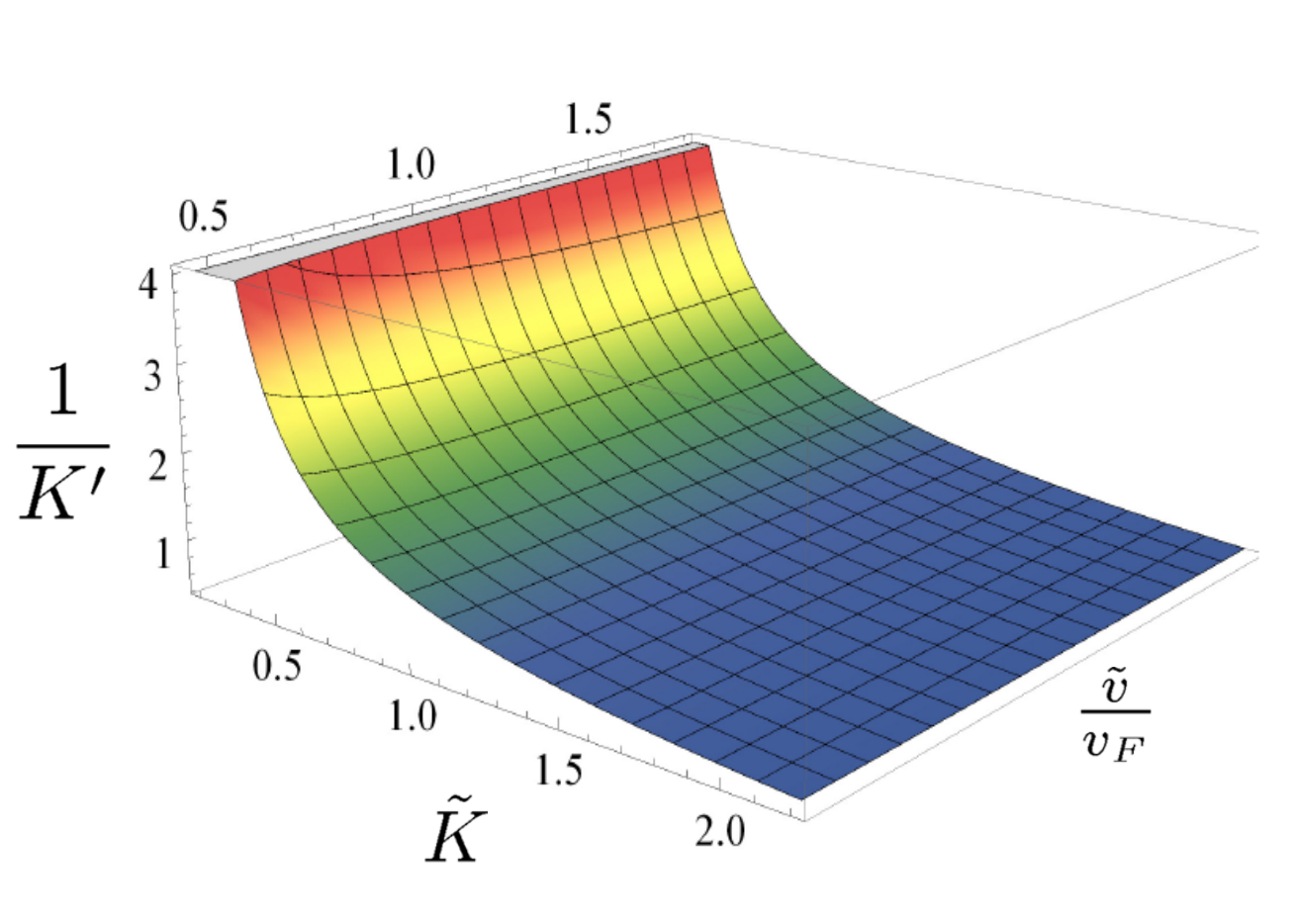}
\caption{ Dependence of $K'$ on the flow parameters $\tilde K$ and $\tilde v$. Here we used $K_w=0.8$, $K_{\rho}=0.2$ and $v=0.5v_F$. The function $K'$ is a monotonically increasing function of $\tilde K$.}\label{fig:Kprime}
\end{figure}

We now compute the RG equations for this model at one-loop level. Using Eqs.\eqref{eqn:OPE} and \eqref{eq:corr_fun},
we find
\begin{align}
\label{eqn:K-v-flow}
\frac{d{}}{d\ell}\Bigl({\tilde K}{\tilde v}\Bigr) &=
\sqrt{v v_F} {f_2}\!\left(\mbox{$\frac{v_F}{v}$}\right)\,y^2,\\
\frac{d{}}{d\ell}\Bigl({\tilde K}{\tilde v}^{-1}\Bigr) &=
\frac{1}{\sqrt{v v_F}} {f_0}\!\left(\mbox{$\frac{v_F}{v}$}\right)\,y^2,
\end{align}
where $y=\Delta_P a/\sqrt{v v_F}$ and the dimensionless function
\begin{equation}
{f_n}(\kappa) = \frac{1}{8\pi}\int_{-\infty}^\infty dz \frac{z^n}{(\kappa^{-1}+{z^2})^{\frac{1}{2K_\rho}}
(\kappa+{z^2})^{\frac{1}{K_w}}}.
\end{equation}
As follows from Eqs.\eqref{eqn:K-v-flow}, the RG equations for $\tilde K$ and $\tilde v$ are given by
\begin{align}
\label{eqn:K-flow_a}
\frac{d\tilde K}{d\ell} = \frac{y^2}{2}
\left[ \frac{\tilde v}{\sqrt{v v_F}}f_0\!\left(\frac{v_F}{v}\right)
+ \frac{\sqrt{v v_F}}{\tilde v}f_2\!\left(\frac{v_F}{v}\right) \right]\\
\frac{d{\tilde v}}{d\ell}=\frac{1}{2\tilde K}\left[
\sqrt{v v_F}f_2\!\left(\frac{v_F}{v}\right)-\frac{\tilde v^2}{\sqrt{v v_F}}f_0\!\left(\frac{v_F}{v}\right) \right]\label{eqn:K-flow_b}.
\end{align}
Thus, according to these Kosterlitz-Thouless-type RG equations,
Eqs.~(\ref{eqn:y-flow2}),(\ref{eqn:K-flow_a}) and \ref{eqn:K-flow_b}, we see that
${\tilde K}$ grows. The parameter $K'$ has a complicated dependence on
${\tilde K}$ and $\tilde v$ which follows from Eq.~\eqref{eq:corr_fun}. However, at small initial velocity mismatch $|\delta v|\equiv |v-v_F|\ll v_F$, one finds
\begin{align}
\left(K'\right)^{-1}&\approx\left[\tilde K-\frac{(2K_{\rho}-K_w)^2}{4(2K_{\rho}+K_w)}\right]^{-1}\\
&-\frac{32 \tilde K(K_w^2-4K_{\rho}^2)K_{\rho}(3K_w+2K_{\rho})}{\left[4(2K_{\rho}+K_w)\tilde K-(2K_{\rho}-K_w)^2\right]^3} \frac{\delta v^2}{v_F^2}\nonumber
\end{align}
One can see that the $\delta v^2$-correction is quickly decaying with $\tilde K$ and thus do not change qualitatively our results obtained for the $v=v_F$ case, cf. with Eq.~\eqref{eq:Kprimeeq_a}. In general, we find that $K'$ is a monotonically increasing function of $\tilde K$, see Fig.\ref{fig:Kprime}. Thus, the growth of ${\tilde K}$ implies the growth of $K'$. Therefore,
once relevant, $y$ will grow to strong coupling $y(l^*)\sim 1$ and pin $\theta_-$.
In this case, $\theta_-$ drops out from Eq.~\eqref{fullaction3} and the effective action now reads:
\begin{align}
\label{fullaction-velocities-+only}
\!\!S
\!=\!\frac{1}{2\pi}\!\int\! dx \,d\tau\,{\tilde K(l^{*})}\!\left( {\tilde v(l^{*})} (\partial_x {\theta_+})^2
 \!+\!  {\tilde v(l^{*})}^{-1}
(\partial_\tau {\theta_+})^2\right).
\end{align}
Following the argument that we used for $v={v_F}$, we observe that
quasi-long-ranged order is stable so long as flux $hc/e$ vortex tunneling
is irrelevant, i.e. when $\sin(4{\phi_+})$ is irrelevant. This occurs when
$4{\tilde K}>1$. Note that this can be satisfied even if ${K_\rho}<1$
and ${K_w}<1$. As in the equal velocity case, two systems, neither of which could sustain superconductivity on their own, can develop superconduct when in proximity as a result of the helical nature of one of the systems.

We thereby arrive at the
model of Ref. \onlinecite{QMajorana}: a 1D
channel that is proximity-coupled to a quasi-long-range-order superconducting
wire. As shown there, such a wire supports Majorana zero modes. We also expect our results discussed in this section to apply to multichannel nanowires with an odd number of occupied subbands coupled to superconducting droplets, see, e.g., Refs.~\onlinecite{Wimmer'10, Potter2010, Lutchyn2011}.

\section{Discussion}
\label{sec:discussion}

In this paper, we have adopted the point of view that SrTiO$_3$
has the seeds of both magnetism and superconductivity. However,
these local tendencies only come to fruition when brought into contact with
a metallic layer or 1D channel. We have focussed on the latter case, for
reasons of tractability as well as potential relevance to the experiments
of Refs. \onlinecite{Levy1, Levy2}, but we believe that our general mechanism
works in 2D as well. We have shown that local moments in SrTiO$_3$
that are near the LAO/STO interface can order ferromagnetically, as a result
of their interaction with mobile electrons at the interface. We have also shown
that droplets of local superconductivity in STO -- which would interact too weakly
to develop superconducting order if left to their own devices -- can develop
superconducting order as a result of their interaction with mobile electrons at the interface.
Finally, we have noted that the interface electrons can form a topological superconducting
state as a result of their proximity to ferromagnetic and superconducting order.

We have shown that our model leads to ferromagnetism by
two different calculations: a large-$N$ calculation and a DMRG calculation.
Both calculations find a ferromagnetic state with spins pointing in the plane, {\it i.e.} along $y$-axis with $x$ being the direction along the wire.
(In the DMRG calculation, the polarization may be either partial or full, depending
on the strength of the spin-orbit coupling.) Interestingly, our large-$N$ calculation
finds a ferromagnetic state even at weak coupling, which is a feature of the
band structure in the presence of Rashba spin-orbit coupling.
A 1D wire with sufficiently strong spin-orbit coupling and Zeeman
field will form a helical wire. Our calculations -- both
large-$N$ and DMRG -- show that our model
gives rise to a helical wire. Remarkably, recent transport measurements
can be interpreted as evidence that 1D channels at the LAO/STO interface
are helical wires \cite{Cheng12}.

In a helical wire, it is possible for $s$-wave superconductivity to
coexist with a magnetic moment. As a result our model allows for
a proximity coupling between $s$-wave superconducting droplets
in STO and ferromagnetic electrons at the LAO/STO interface.
We have analyzed our model for superconductivity by mapping it to
the theory of a single 1D boson in the vicinity of the Kosterlitz-Thouless transition.
We find that such a model could be on the disordered side of the
Kosterlitz-Thouless transition for very weak coupling between the droplets
and a 1D channel at the interface but it could be on the ordered side of the
transition if the coupling is sufficiently strong.

Our results on magnetism and superconductivity imply that
the superconducting state of a 1D channel at the LAO/STO interface
is in a topological superconducting state. This actually stabilizes
the system against quantum phase slips: $2\pi$ phase slips are forbidden,
and only $4\pi$ phase slips, which are less relevant in the RG sense,
could disrupt the superconductivity. Furthermore, this topological
superconducting state supports Majorana fermion zero modes,
whose presence would lead to a $4\pi$-periodic ac Josephson effect~\cite{1DwiresKitaev, 1DwiresLutchyn}.

\begin{figure}[tbp]
\centerline{\includegraphics[width=3.4in]{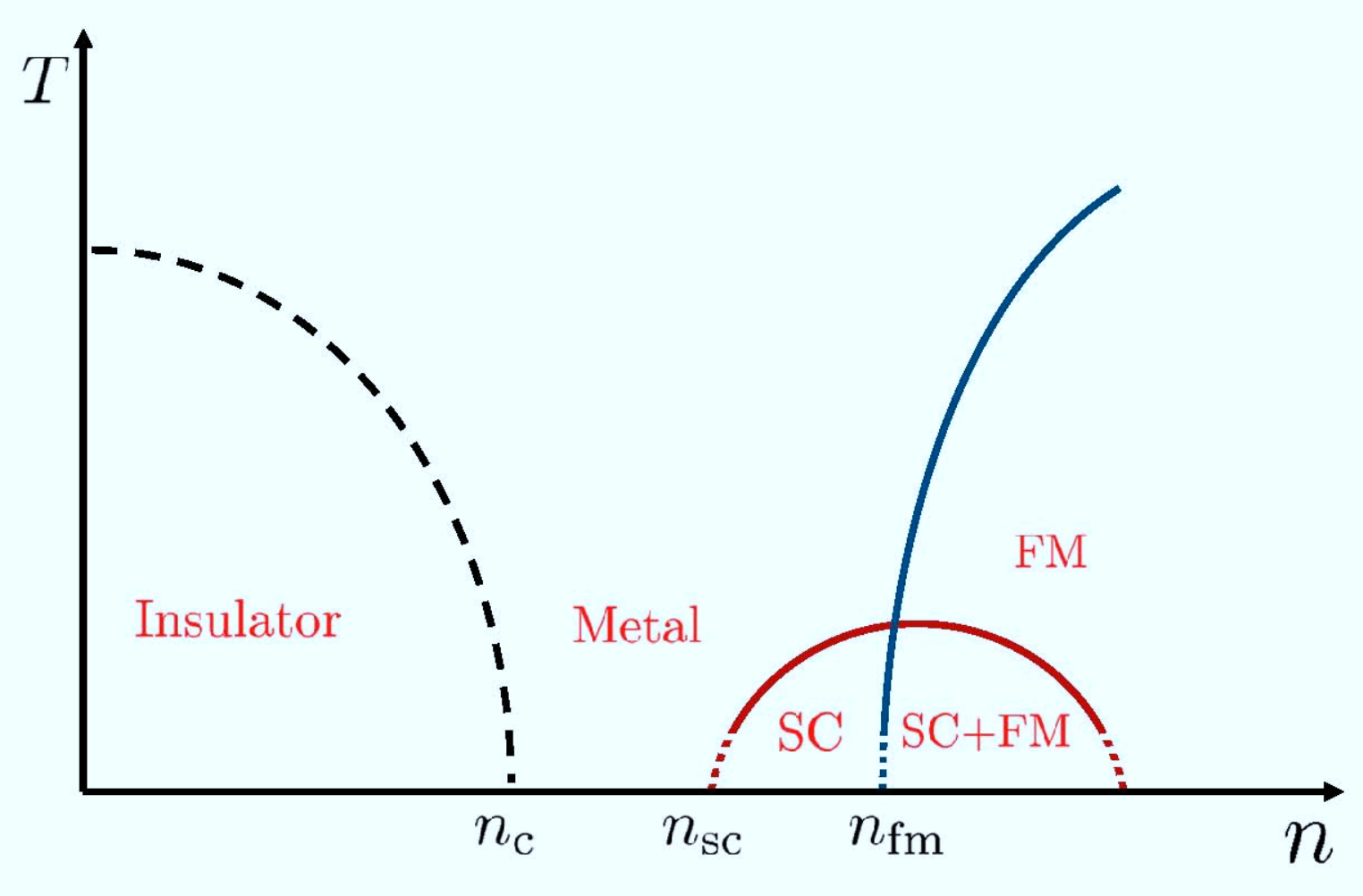}}
\caption{A schematic phase diagram for the LAO/STO interface
as a function of carrier concentration.}
\label{Fig:phase-diagram}
\end{figure}

The main goal of our work has been to show why magnetism
occurs, why superconductivity occurs, and why they can coexist.
However, magnetism is probably not found at all carrier concentrations
above the metal-insulator transition \cite{Thiel06}. Neither is superconductivity.
A plausible schematic phase diagram is given in Fig. \ref{Fig:phase-diagram},
based on the phase diagram in Refs. \onlinecite{Stemmer,Jeremy-email}.
Therefore, it is also important to understand when and why they do not occur.
If the magnetic order is too strong, so that the spins are fully polarized,
then our mechanism does not work. This may explain why superconductivity
is suppressed at large carrier concentration. However, a more detailed
understanding of the phase diagram is definitely an important target for
further investigation.

We note that our calculations rely heavily on simplifying features
of one dimensional systems -- the applicability of DMRG calculations to
the magnetic ordering of the system and the applicability of bosonization
to the superconducting ordering of the system. It would be interesting to
give a fully two-dimensional analysis of a model similar to ours.  Furthermore, it would be useful to investigate other related materials exhibiting the same phenomena.  In particular, recent work on epitaxially grown GdTi${\rm O}_3$-SrTi${\rm O}_3$ interfaces \cite{Stemmer} indicates that ferromagnetism and superconductivity can also coexist in such systems.  Being much cleaner than LAO/STO from a materials point of view, these interfaces might provide an attractive environment for the investigation of the ideas proposed in this paper.

\acknowledgements We would like to thank Guanglei Cheng, Harold Hwang,
Jeremy Levy, Susanne Stemmer, and Joshua Veazey for discussions. H.C.J. is
partially supported by the the KITP NSF grant PHY05-51164 and the
NSF MRSEC Program under Award No. DMR 1121053. C.N. is supported by
the DARPA QuEST program and the AFOSR under grant FA9550-10-1-0524.
We thank the Aspen Center for Physics for hospitality and support
under NSF grant \#1066293.


%

\end{document}